\UseRawInputEncoding
\documentclass[a4paper,fleqn,usenatbib]{mnras}

\usepackage{subfigure}
\usepackage{epsfig}
\usepackage{rotating}
\usepackage[T1]{fontenc}
\usepackage{ae,aecompl}
\usepackage{tikz}

\usepackage{graphicx}	
\usepackage{amsmath}	
\usepackage{amssymb}	

\usepackage{color}
\usepackage{natbib}

\usepackage{booktabs}

\newcommand{\Ms}{{\ensuremath{\mathrm{M}_{\odot} }}}

\title[Radio Emission from GHZ9]{Radio Signatures of a Massive Black Hole in GHZ9 at z $\sim$ 10}
\author[Latif and Whalen]{Muhammad A. Latif$^{1}$\thanks{E-mail: latifne@gmail.com}, 
Daniel J. Whalen$^{2}$
\\
\\
$^{1}$ Physics Department, College of Science, United Arab Emirates University, PO Box 15551, Al-Ain, UAE \\ 
$^{2}$Institute of Cosmology and Gravitation, University of Portsmouth, Portsmouth PO1 3FX, UK  \\
}

\date{Accepted XXX. Received YYY; in original form ZZZ}

\pubyear{2024}

\begin{document}
\label{firstpage}
\pagerange{\pageref{firstpage}--\pageref{lastpage}}
\maketitle

\begin{abstract}

Synergies between the {\em James Webb Space Telescope} ({\em JWST}) and the {\em Chandra} X-ray observatory have advanced the observational frontier by detecting a handful of active galactic nuclei (AGNs) beyond $z \sim$ 10. In particular, the recent discovery of a candidate $\rm 8 \times 10^7$ \Ms\ black hole (BH) in the galaxy GHZ9 at $z =$ 10.4 favors massive seed formation channels for these objects.  Motivated by prospects for their detection in radio by recent studies, we estimate radio fluxes for GHZ9 and explore the possibility of their detection with the Square Kilometer Array (SKA) and next-generation Very Large Array (ngVLA). We find that ngVLA should be able to detect radio emission from GHZ9 for integration times as short as 1 hr while SKA will require integration times of up to 100 hr. We also find that radio emission from the BH can be distinguished from that due to H II regions and supernovae in its host galaxy. The detection of a few hundred nJy radio signal at frequencies $> 2$ GHz will be a smoking gun for the presence of a BH in GHZ9.
\end{abstract}


\section{Introduction}

Supermassive black holes (SMBHs) of a few million to billion solar masses reside not only at the centers of galaxies today but have now been discovered just a few hundred Myr after the Big Bang (\citealt{yang20,wang21} -- see \citealt{fbs23} for a recent review).  {\em JWST} has now revealed a population of previously hidden low-luminosity AGNs at $z >$ 4, with a handful of candidates at $z>$ 10 \citep{Kok23,Lamb23,Lars23,Koc23,Maio23,Maio23a,furt24}. These BHs are typically a few $\rm 10^5-10^7$ \Ms\ with bolometric luminosities of $\rm 10^{44}- 10^{45}$ erg s$^{-1}$.  This suggests that massive seeds from the direct collapse scenario might be their origin \citep{wf12,latif13c,hos13,tyr17,hle18b,latif22b,Sch22,Vol23,pat23a}. Synergies between {\em Chandra} X-ray and {\em JWST} near infrared (NIR) observations have now revealed BHs beyond $z \sim$ 10, like the $\rm 4 \times 10^7$ \Ms\ BH in UHZ1 at $z =$ 10.3 \citep{Bod23,Gould23} and now the $\rm 8 \times 10^7$ \Ms\  BH in GHZ9 at $z =$ 10.4 \citep[][hereafter K24]{Kov24}.

GHZ9 was originally found in the UNCOVER and GLASS surveys by \citet{Atek23} and \citet{Cast23} as a lensed dropout Lyman-break galaxy behind the Abell 2744 galaxy cluster. Its redshift $z$, stellar mass, $M_{\star}$, and star formation rate (SFR) are derived from stellar energy distribution (SED) models and vary from $\rm 10.37^{+0.32}_{-1.09}$ - $9.35^{+0.77}_{-0.35}$, $\rm 4.9 ^{+2.0}_{-2.6} \times 10^7$ - $\rm 3.3^{+2.9}_{-2.4} \times 10^8$ \Ms, and $\rm 0.56 ^{+0.23}_{-0.29}$ - $\rm 14.4^{+15.0}_{-7.3}$ \Ms\ yr$^{-1}$, respectively. Recently, GHZ9 was confirmed to be an AGN candidate in Chandra X-ray observations (K24). These BHs are overmassive relative to their host galaxies, 2-3 orders of magnitude above the local $\rm M_{\rm BH} - M_{\star}$ relation. Both NIR and X-rays are obscured by dust in the host galaxies while radio waves escape.  Therefore, radio followups of these sources will be crucial in confirming the presence of BHs in future obscured AGN candidates and probing their formation pathways.

Previous studies indicate that BHs $\gtrsim 10^6$ \Ms\ could be detected at $z \lesssim 13 - 14$ by ngVLA and SKA in the coming decade \citep{W23,L24b,L24}. Recently, \cite{W23} found that SKA and the ngVLA could detect radio emission from the BH in UHZ1 with integration times of 10 - 100 hr and 1 - 10 hr, respectively.  Here we determine if the existence of a BH in GHZ9 can be confirmed by detection of radio flux, taking into account thermal bremsstrahlung emission from H II regions and synchrotron flux from SN remnants in the host galaxy.  We describe our flux estimates in Section 2, present flux densities at 0.1 - 10 GHz in Section 3 and conclude in Section 4.

\section{Numerical Method}

\label{sec:method}

\subsection{Radio Flux from BH}

We estimate radio flux from the BH in GHZ9 with fundamental planes (FPs) of BH accretion at 0.1 - 10 GHz. FPs are empirical relations between $M_\mathrm{BH}$, the nuclear radio luminosity at 5 GHz, $L_\mathrm{R}$, and the nuclear X-ray luminosity at 2 - 10 keV, $L_\mathrm{X}$ \citep{merl03}. They span over five orders of magnitude in BH mass and are valid down to $M_\mathrm{BH} \sim 10^5$ \Ms\ \citep{gul14}.  In order to compute the radio flux in the observer frame, we first calculate $L_\mathrm{X}$ in the source frame from the bolometric luminosity, $L_{\mathrm{bol}}$, with Equation 21 of \citet{marc04}:
\begin{equation}
\mathrm{log}\left(\frac{L_\mathrm{bol}}{L_\mathrm{X}}\right) = 1.54 + 0.24 \mathcal{L} + 0.012 \mathcal{L}^2 - 0.0015 \mathcal{L}^3,
\end{equation}
where $L_\mathrm{bol}$ is in units of solar luminosity and $\mathcal{L} = \mathrm{log} \, L_\mathrm{bol} - 12$. For GHZ9, $M_\mathrm{BH}$ is $\rm 8 \times 10^7$ \Ms\ and $L_{\mathrm{bol}}$ is $\rm 1.0 \times 10^{46}$ erg s$^{-1}$ (K24). We then find $L_\mathrm{R}$ from $L_\mathrm{X}$ from the FPs:
\begin{equation}
\mathrm{log} \, L_\mathrm{R} (\mathrm{erg \, s^{-1}})= \alpha \, \mathrm{log} \, L_\mathrm{X} (\mathrm{erg \, s^{-1}}) + \beta \, \mathrm{log} \, M_\mathrm{BH} (\mathrm{M}_{\odot})+ \gamma,
\end{equation}
where $\alpha$, $\beta$, and $\gamma$ are taken from \citet[][MER03]{merl03}, \citet[][KOR06]{kord06}, \citet[][GUL09]{gul09}, \citet[][PLT12]{plot12}, and \citet[][BON13]{bonchi13} and listed in Table 2 of \cite{L24b}. We also consider the FP of \citet[][GUL19]{gul19} in their Equation 19, which has the following form:
\begin{equation}
R \, = \, -0.62 + 0.70 \, X + 0.74 \, \mu,
\end{equation}
where $R =$ log($L_\mathrm{R}/10^{38}$ erg s$^{-1}$), $X =$ log($L_\mathrm{X}/10^{40}$ erg s$^{-1}$) and $\mu =$ log($M_\mathrm{BH}/10^{8}$ \Ms). 

Radio flux from GHZ9 in a given band in the observer frame today does not originate from 5 GHz in the source frame because of cosmological redshifting.  Flux that is redshifted into ngVLA and SKA bands today is determined from $L_\mathrm{R} =$ $\nu L_{\nu}$ in the source frame, assuming that $L_{\nu} \propto \nu^{-\alpha}$. Observations of radio sources in the local universe show that the median value of $\alpha$ is 0.7 \citep{ccb02} while radio emission by quasars at $z >$ 5 exhibit a median value of 0.3 \citep{glou21}. Both studies suggest an intrinsic scatter in the value of $\alpha$ for individual sources. To cover a reasonable range of spectral profiles, we consider $\alpha =$ 0.7 and 0.3. We calculate the spectral flux at $\nu$ in the observer frame from the spectral luminosity at $\nu'$ in the rest frame from
\begin{equation}
F_\nu = \frac{L_{\nu'}(1 + z)}{4 \pi {d_\mathrm L}^2},
\label{eq:flx}
\end{equation}
where $d_\mathrm L$ is the luminosity distance and $\nu' = (1+z) \nu$.  We use cosmological parameters from \textit{Planck}:  $\Omega_{\mathrm M} = 0.308$, $\Omega_\Lambda = 0.691$, $\Omega_{\mathrm b}h^2 = 0.0223$, $\sigma_8 =$ 0.816, $h = $ 0.677 and $n =$ 0.968 \citep{planck2} and boost the flux by the lensing factor $\mu=1.26$ \citep{Atek23}.

\subsection{Radio Emission from H II Regions}

Thermal bremsstrahlung in H II regions due to massive star formation also produces continuum radio emission whose spectral density can be estimated from 
\begin{equation}
L_{\nu} \, \lesssim \, \left(\frac{Q_{\mathrm{Lyc}}}{6.3 \times 10^{52} \, \mathrm{s^{-1}}}\right) \left(\frac{T_{\mathrm{e}}}{10^4 \mathrm{K}}\right)^{0.45} \left(\frac{\nu}{\mathrm{GHz}}\right)^{-0.1}
\end{equation}
in units of 10$^{20}$ W Hz$^{-1}$ \citep{con92}, where $Q_{\mathrm{Lyc}} =$ SFR (\Ms \, yr$^{-1}$) $/$ $1.0 \times 10^{-53}$ \citep{ken98} and $T_{\mathrm{e}} =$ 10$^4$ K. For GHZ9, the spectral energy distribution models yield two star formation rates (SFRs), 0.5 and 14.4 \Ms\ yr$^{-1}$ for $z =$ 10.4 and 9.4, respectively (K24). We consider both SFRs for radio emission from H II regions in the host galaxy. 

\subsection{Radio Emission from Supernova Remnants}

Young SN remnants in the host galaxy emit synchrotron radiation that contributes to the total radio emission from AGN, particularly at lower frequencies.  We employ the relation in \citet{con92} to estimate non-thermal radio emission, $L_{\mathrm N}$, from SNe remnants:
\begin{equation}
\left(\frac{L_{\mathrm{N}}}{\mathrm{W \, Hz^{-1}}}\right) \, \sim \, 5.3 \times 10^{21} \left(\frac{\nu}{\mathrm{GHz}}\right)^{-\beta} \left[\frac{\mathrm{SFR}(M > 5 \, M_{\odot})}{M_{\odot} \, \mathrm{yr^{-1}}} \right],
\label{eq:SNe}
\end{equation}
where $\beta \sim$ 0.8 is the nonthermal spectral index and we use the SFRs for GHZ9 mentioned above.  Equation~\ref{eq:SNe} may overestimate the flux because SN radio flux from high-$z$ SNe \citep{wet08a,mw12} in the diffuse H II regions \citep{wan04,ket04} of high-redshift halos \citep{latif22a} are expected to be of the order of nJy.  We assume for simplicity that all stars $>$ 5 \Ms\ produce this flux so our estimates should be taken as upper limits.  We redshift H II region and SNR radio fluxes into the observer frame with Equation~\ref{eq:flx}. As with the BH fluxes, we scale the two fluxes by the lensing magnification factor of 1.26. 

\section{Results and Discussion} 

We compare our results with SKA1 and ngVLA sensitivity limits  for integration times of 1 hr, 10 hr and 100 hr. The SKA rms limits for each frequency band are listed in Table 3 of \cite{B19}  and the ngVLA 5-sigma rms values are taken from \citet{pr18}. The SKA limits are based on SKA1-Low and SKA1-Mid. SKA1-Low will have 131,072 dipole antennae grouped into 512 stations, each with 256 antennae. They will cover 50-350 MHz with a baseline of 65 km. SKA1-Mid will consist of 133 dishes 15-m in radius and, in combination with 64 existing MeerKAT dishes (13.5-m each), will cover 350 MHz to 15 GHz for baselines of up to 150 km. The proposed ngVLA reference design (Long Base Array) will have 244 18-m diameter dishes with a baseline up to  $\sim$ 8000 km.

\begin{figure*}
\begin{center}
\includegraphics[scale=0.54]{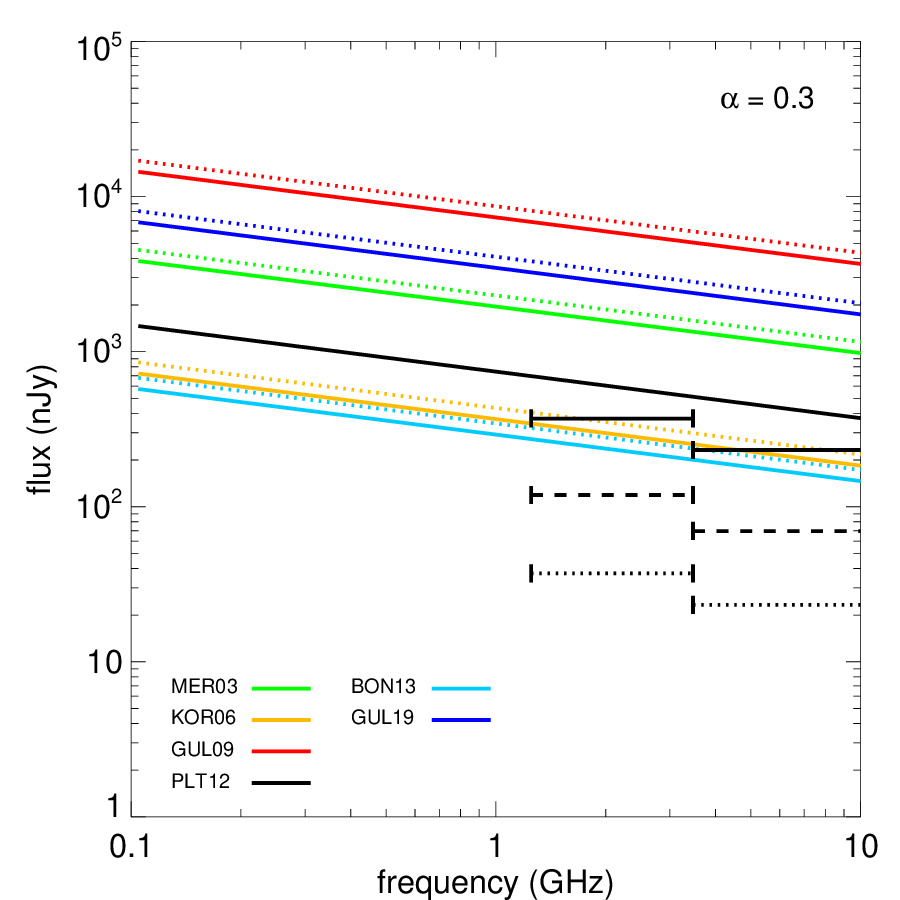}
\includegraphics[scale=0.54]{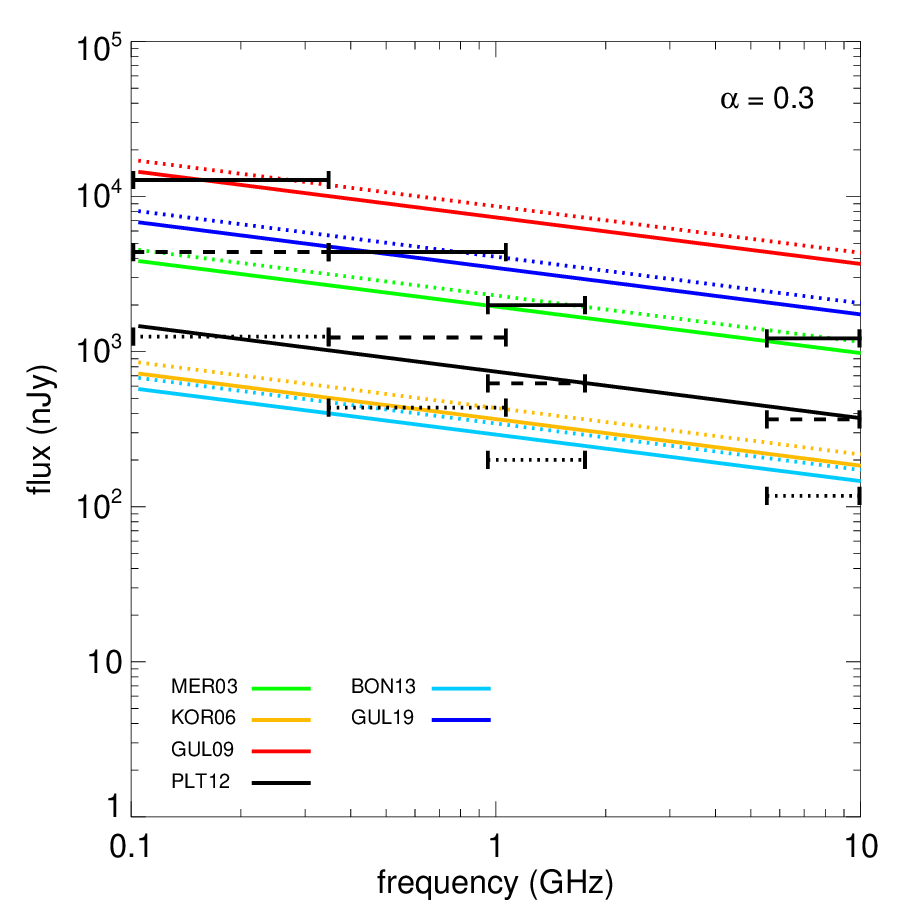}
\includegraphics[scale=0.54]{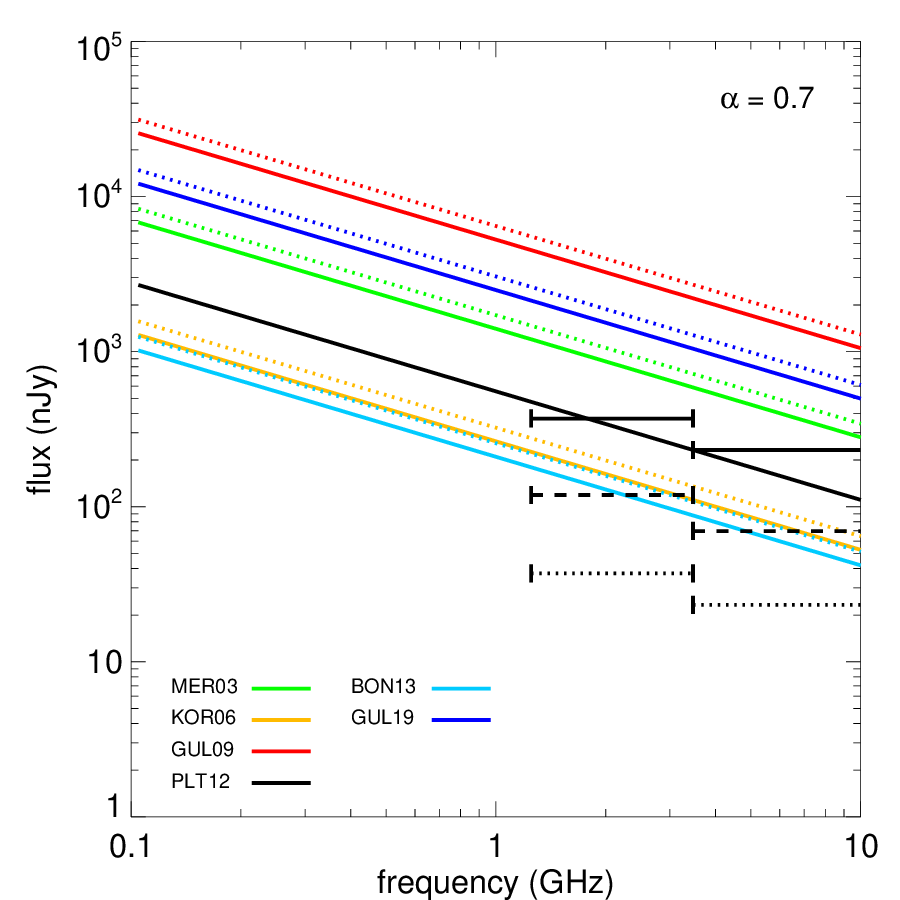}
\includegraphics[scale=0.54]{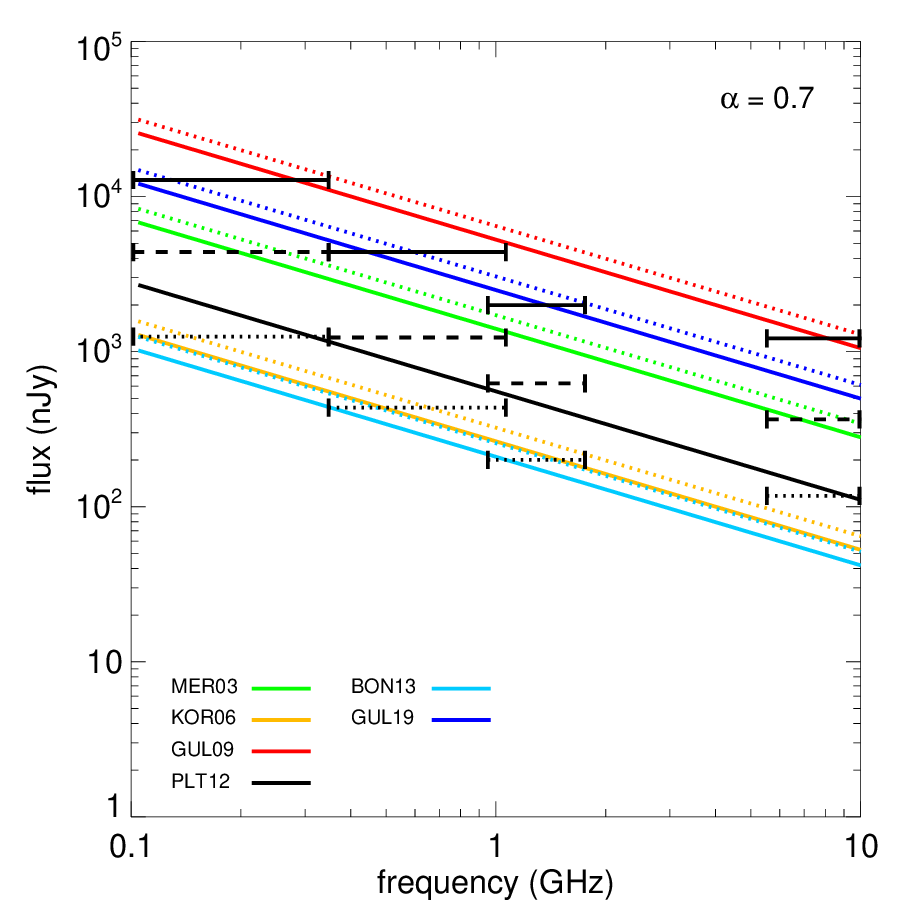}
\end{center}
\caption{Radio fluxes for $z =$ 10.4 (solid lines) and $z =$ 9.4 (dotted lines) are shown here for $\alpha = 0.3$ and $\alpha = 0.7$. Each color represents a FP as shown in the legend. The horizontal lines show detection limits of ngVLA and SKA for integration times of 1 hr (solid), 10 hr (dashed) and 100 hr (dotted), respectively.  The left column for the ngVLA and the right column for the SKA observational limits for different integration times.}
\label{fig:BH}
\end{figure*}

Our BH radio fluxes are shown for $\alpha =$ 0.3 and 0.7 in Figure~\ref{fig:BH}.  They vary from about 500 nJy to 30 $\mu$Jy at 0.1 GHz and 130 nJy to 4.5 $\mu$Jy at 10 GHz for $\alpha =0.3$.  These fluxes do not depend on the range of redshifts inferred for GHZ9, as they differ by a factor of less than two for $z =$ 9.4 - 10.4.  We find that ngVLA should be able to detect GHZ9 radio emission with integration times of 1 hr at $>$ 1 GHz for $\alpha =0.3$ while SKA would require about 100 hr. For $\alpha =0.7$, the BH radio fluxes are about factor of two higher at 0.1 GHz and three times lower at 10 GHz compared to $\alpha =0.3$.  These lower fluxes will require longer integration times of at least 10 hr with ngVLA and 100 hr for SKA. K24 found that the statistical uncertainties in the BH mass and the bolometric luminosity are $8.0^{+3.7}_{-3.2} \times 10^7$ \Ms\  and $1.0^{+0.5}_{-0.4} \times 10^{46}$ erg s$^{-1}$, respectively. We  investigated the potential impact of such uncertainties by considering the lowest BH mass of  $\rm 4.8 \times 10^7$ \Ms\ and bolometric luminosity of $\rm 6 \times 10^{45}$ erg s$^{-1}$ and found that they decrease BH radio flux at most by a factor of 3.  Considering the upper limit in BH mass and $\rm L_{bol}$ would further increase fluxes by a factor of 2 - 3.  We further note that error bars in the FP coefficients may increase fluxes by a factor of few or lower them up to by an order of magnitude.

The maximum flux from H II regions is about 20 nJy at $z =$ 9.4 and below 1 nJy at $z =$ 10.4.  SN radio flux is highest at 0.1 GHz and strongly depends on redshift, with maximums of 30 nJy and 900 nJy at $z =$ 10.4 and 9.4, respectively.  We find that emission from H II regions is too weak to be detected by ngVLA or SKA and that only ngVLA can detect SN flux at $>$ 2 GHz, with integration times of about 100 hr.

\begin{figure*}
\begin{center}
\includegraphics[scale=0.54]{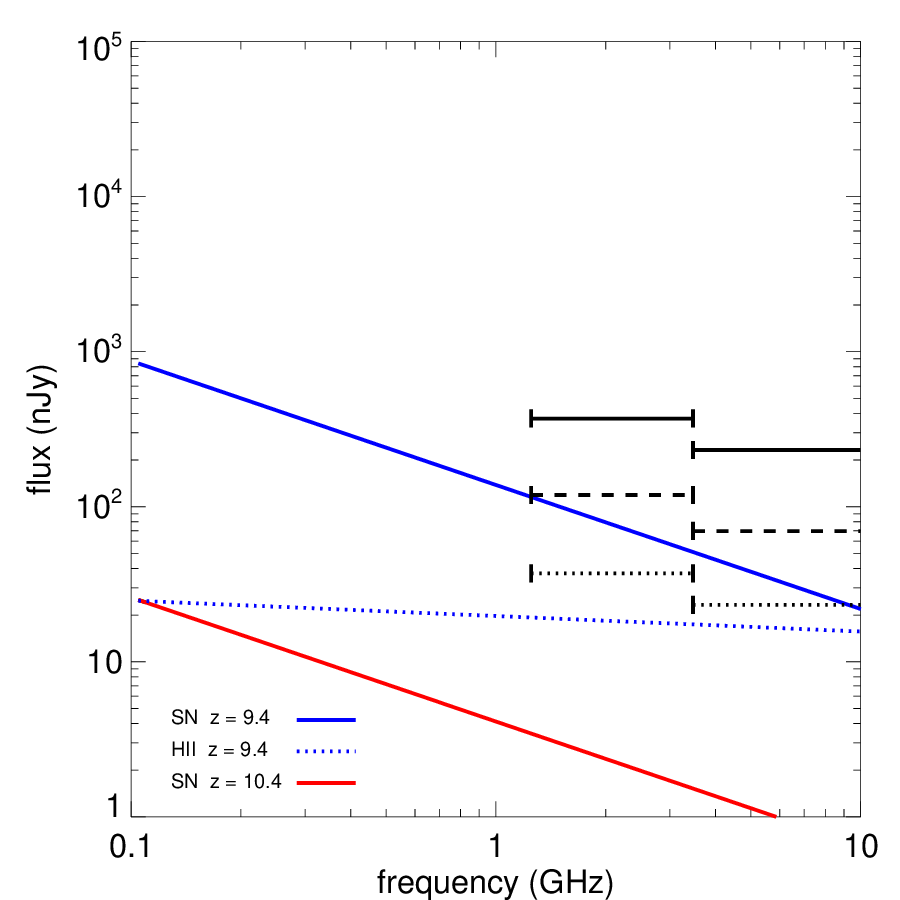}
\includegraphics[scale=0.54]{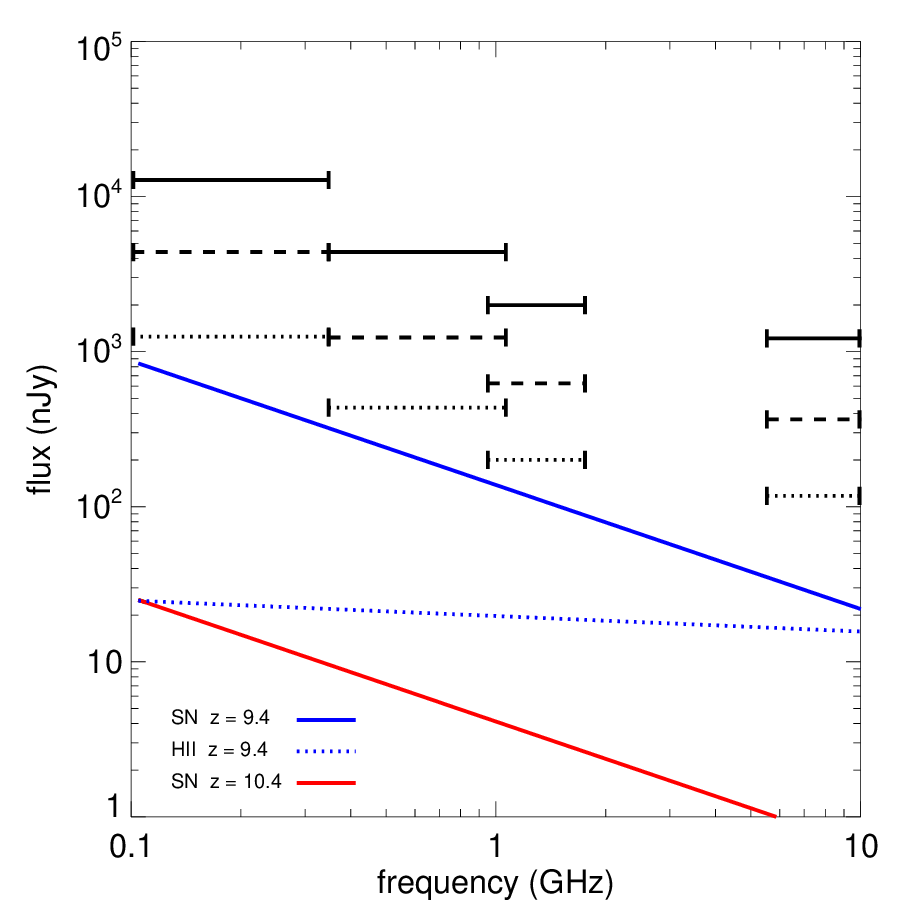}
\end{center}
\caption{Radio emission from H II regions and SN remnants. H II region fluxes for z = 10.4 are below 1 nJy because of the low SFR and are not shown here.The left column for the ngVLA and the right column for the SKA show detection limits for different integration times of 1 hr (solid), 10 hr (dashed) and 100 hr (dotted), respectively.}
\label{fig:HII}
\end{figure*}

K24 derived the bolometric luminosity from $L_{\rm X}$ at 0.5 - 3 KeV using the correction factor from \cite{L12}. Our results are subject to uncertainties in the bolometric correction factor although \cite{D20} found that it has no dependence on the source redshift.  The relation between $L_{\rm R}$ and $L_{\rm X}$ in the FPs varies from $\rm L_{\rm R} \propto L_{\rm X}^{0.39 - 0.71}$ which is close to observed range for radio quiet AGN \citep{Li08}. Therefore, our estimates will be applicable for the case of radio quiet candidate AGN in GHZ9. In the case of radio loud candidate AGN in GHZ9, we expect the flux to be even higher than reported here. Similarly, the uncertainty in the redshift suggest it to be as high as 11.5, this would not change our conclusions as the differences in the BH flux are expected to be within a factor of 1.5. Recently, \cite{napo24} have spectroscopically confirmed that its redshift is 10.145 and  BH mass $1.6 \pm 0.31 (0.48 \pm 0.09) \times 10^8$  \Ms\ depending on the adapted spectral index $\rm \Gamma =$ 2.3(1.8). This new redshift and BH mass fall within the ranges we have considered here.

We assume here radio emission primarily arises from the AGN core, which may include some emission from the base of compact jets, and we have ignored radio signals due to synchrotron emission by extended jet.  Compact jets from high-redshift quasars have only been observed in one or two cases and just on scales of a kpc or less \citep{mom18,con21}.  Jets are not expected here because they have generally only been observed at $L_{\mathrm{bol}} \lesssim 0.01 \, L_{\mathrm{Edd}}$ or $L_{\mathrm{bol}} \gtrsim L_{\mathrm{Edd}}$ and the BH in GHZ9 is accreting at a fraction of the Eddington limit.  Even if a jet did form, its radio emission would likely be quenched because relativistic electrons preferentially upscatter cosmic microwave background (CMB) photons in energy instead of emitting synchrotron flux at $z >$ 6, a phenomenon known as CMB muting \citep{gh14,fg14}. CMB quenching has no effect on the FP fluxes because they originate from the nuclear region of the AGN.

Our estimates show that radio flux from GHZ9 could be detected by upcoming radio observatories such as ngVLA and SKA.  Although GH9 remains undetected in the current VLA surveys (such as NVSS and VLASS) it could be detected with VLA with longer integration times. VLA bands L (1.5 GHz), S (3 GHz) and C (6 GHz) are suitable for the detection of GHZ9, which would require integration times of 24 - 100 hours based on  2, 1 and 0.5 $\mu$Jy for 24 hrs \citep[see Figure 1 of][]{smol17}.  BH radio flux from GHZ9 dominates that of its host galaxy, particularly at higher frequencies ($\gtrsim 2$ GHz). Keeping in mind that our SN radio fluxes are an upper limit, any radio emission above 50 nJy at $\ge 8$ GHz would be conclusive evidence of the existence of a BH in GHZ9, the most distant one ever observed. GHZ9 will be an excellent subject for upcoming radio surveys by ngVLA and SKA and will open the first synergies between JWST, X-ray and radio observations.

\section{acknowledgments}

MAL thanks the UAEU for funding via SUREPlus grant No. 12S182.  

\section{Data Availability Statement}

The data in this study will be made available upon reasonable request to the corresponding author.

\bibliographystyle{mnras}
\bibliography{ref.bib}

\end{document}